\documentclass[preprint,12pt]{elsarticle}

\usepackage{amssymb}
\usepackage{amsmath}
\usepackage{graphicx}
\usepackage{multirow}
\usepackage{subcaption}
\usepackage{hyperref}
\usepackage{cleveref}
\usepackage{placeins}
\usepackage{siunitx}

\journal{Nuclear Physics B}

\begin{document}

\begin{frontmatter}


\title{Direct stroke measurement of Piezos for  cavity frequency tuner of the ILC prototype cryomodule using a Laser Displacement Sensor}
\author[a]{Rishabh Bajpai\corref{cor1}}
\ead{bajpai@post.kek.jp}
\author[a,b]{Mathieu Omet}
\author[a,b]{Ashish Kumar}
\cortext[cor1]{Corresponding author}
\address[a]{High Energy Accelerator Research Organization,
	1-1 Oho, Tsukuba, Ibaraki 305-0801, Japan}
\address[b]{The Graduate University for Advanced Studies, Department of Accelerator Science, School of High Energy Accelerator Science, High Energy Accelerator Research Organization (KEK), Tsukuba, Ibaraki 305-0801, Japan}	
\begin{abstract}
Piezoelectric actuators are critical for achieving high accelerating gradients and preventing RF trips in narrow-bandwidth superconducting radio-frequency (SRF) cavities by compensating for detuning caused by Lorentz force detuning. 
Depending on the maximum acceleration gradient an appropriate piezo stroke requirement has to be fulfilled. 
Since the stroke of piezo actuators decreases at cryogenic temperatures, evaluating their performance under such conditions is essential.
Common characterization methods either use the SRF cavity itself as a sensor or rely on capacitance measurements during cool-down. 
Both these approaches do not measure the stroke directly and involve a trade-off between measurement precision and experimental simplicity, as well as cost and time.
We developed a new method for the direct and precise measurement of piezo stroke at cryogenic temperature inside a cryocooler-cooled cryostat using a laser displacement sensor. 
The setup was used to characterize and evaluate two piezo actuators for cavity frequency tuners of the ILC prototype cryomodule, which is currently being built at KEK.
In this article we are reporting on the development, setup, test, and application of this novel method, allowing the direct stroke measurement of piezos in vacuum and at cryogenic temperatures.
\end{abstract}



\begin{keyword}
 Piezo \sep Tuner \sep SRF Cavity \sep ILC \sep Cryogenics \sep Laser Vibrometery 

\end{keyword}

\end{frontmatter}



\section{Introduction}

The International Linear Collider (ILC) is a proposed electron-positron linear collider under consideration for construction in Japan \cite{ilc}.
With the initial center-of-mass energy of 250 GeV, ILC \cite{ilc2501,ilc2502} will act as Higgs Boson factory conducting precise measurement of its properties\cite{higgs} and probing new physics beyond the standard model.

The key components of ILC are the two Linacs, which will accelerate the electron and positron beam to 125GeV each. 
The linacs utilize TESLA technology, with beams accelerated through superconducting RF cavities operating at 2 K.
The ILC cavities are nine-celled structures made of high-purity niobium, and operate at 1.3 GHz. 
Each nine-cell cavity is housed inside a liquid helium tank and combined with auxiliaries (such as fundamental-mode power coupler, mechanical frequency tuner, etc.). 
Eight or nine of these cavity with auxiliaries are housed in a cryogenic vacuum vessel called cryomodule. 
At KEK, a five-year project is ongoing to manufacture\cite{itn}, construct, and test a prototype cryomodule (CM) that satisfies the ILC specifications and is scheduled for cooling and RF test in 2028 without beam acceleration. 

\subsection{Lorentz Force Detuning and Compensation}
\label{subsec1}

SRF cavities are electromagnetic resonators with a high quality factor and very narrow bandwidth. 
These superconducting niobium cavities are made from thin sheets to reduce material cost and to allow efficient heat transfer. 
However, the thin cavity walls make them sensitive to mechanical deformation, which causes a shift in the cavity frequency known as detuning.
When the cavity is driven by RF power under ideal conditions with no detuning, most of the supplied power is used for beam acceleration. 
If the cavity frequency shifts from the nominal value, the cavity and the power coupler are no longer impedance matched. 
As a result, more RF power is required to maintain the desired accelerating gradient because part of the power is reflected and lost.

ILC will be a pulsed machine with a repetition rate of 5Hz.
The Lorentz forces generated due to large time varying magnetic fields and wall currents excite the mechanical modes of the cavity.
This mechanical excitation leads to detuning of the cavity known as Lorentz Force Detuning (LFD) \cite{lfd}.
Since the klystrons are operated near saturation level, compensating for the detuning is critical to achieve the average target acceleration gradient of 31.5 MV/m.

\begin{figure}[b!]
	\centering
	\includegraphics[width=0.65\linewidth]{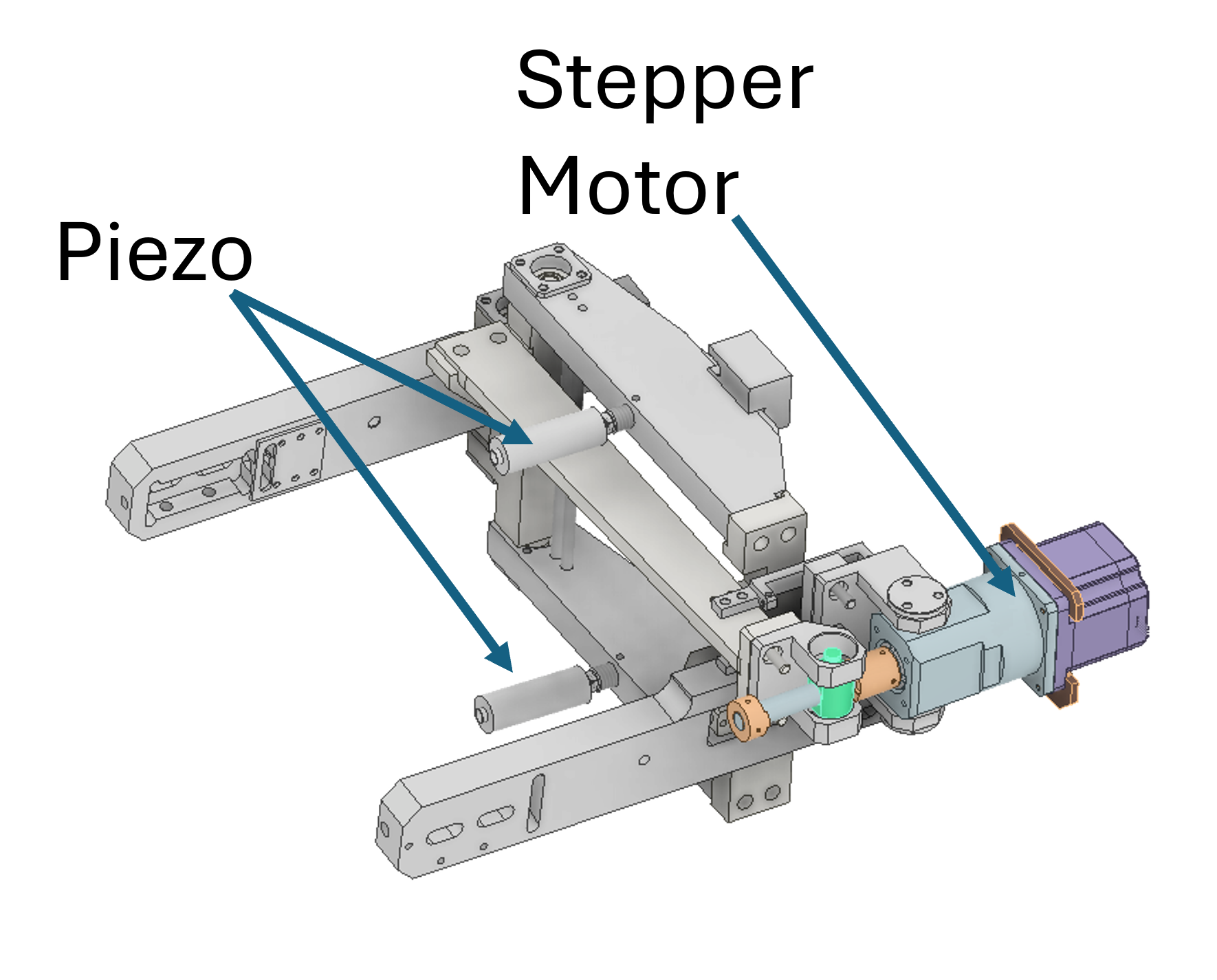}
	\caption{Schematic of the tuner developed for ILC prototype cryomodule, indicating the position of piezos and stepper motor.}
	\label{fig:tuner}
\end{figure}

The compensation for this detuning is done using a mechanical tuner.
The tuner is mounted onto the cavity, and consists of two types of actuators: a slow/coarse motor and fast/fine piezo actuator.
\Cref{fig:tuner} shows the CAD model of the tuner for ILC prototype cryomodule and details of the how the tuner works can be found in \cite{tup58}.
LFD compensation is carried out by a piezo placed at one end of the cavity. 
The piezo pushes on the cavity, changing its length and canceling the detuning caused by Lorentz forces. 
For nine-cell cavities, the typical detuning observed for a 1 mm change in length is about 300 kHz \cite{300Khz-1,300Khz-2}.
A detailed overview of cavity electromagnetic model, detuning and its compensation can be found in \cite{tup60, Qui}.

\subsection{Piezo Requirements}
LFD depends on the square of acceleration gradients in the cavity given by \cref{eq:1}
\begin{equation} 
	\Delta f=-k_{L}E_{acc}^{2} \label{eq:1}
\end{equation}
where,
$k_{L}$ is the LFD coefficient and varies from 0.9 to 1.3 Hz (MV/m)$^{-2}$ for 1.3 GHz TESLA-type cavities and is calculated for the filling and flattop of a pulse. 
This coefficient is calculated using FEM simulations \cite{lfd_coff}.
For the ILC prototype cryomodule cavities, $k_{L}$ (filling and flattop) was estimated to be \textasciitilde1.27.
However, since LFD compensation in the ILC prototype cryomodule will be carried out only during the flattop, $k_{L}$ during the flattop was estimated to be \textasciitilde0.635 and used to determine the piezo requirements.
Furthermore, while the average gradient is 31.5 MV/m, recently some cavities have reached a gradient of 40 MV/m during vertical tests. 
Therefore, the expected detuning is calculated to be 0.63 kHz and 1.016 kHz for 31.5 MV/m and 40 MV/m, respectively. 
This corresponds to cavity length changes of 2.1 \SI{ }{\micro\meter} and 3.38 \SI{ }{\micro\meter} to compensate for LFD at gradients of 31.5 MV/m and 40 MV/m, respectively.
Therefore, the piezo stroke requirement is set to be at least 3.38 \SI{ }{\micro\meter} at 20 K, at which the piezos will be operated \cite{stroke_cavity}.

A common approach to evaluate whether the piezo stroke is sufficient to compensate for Lorentz force detuning (LFD) is to mount the actuator on an SRF cavity and measure the detuning produced when the piezo is driven \cite{stroke_cavity}. 
However, this method requires a fully assembled cavity and a liquid helium cooling system, making it time-consuming and costly, and therefore impractical for routine characterization.
Another approach is to measure the change in piezo capacitance during cool-down and estimate the reduction in stroke by assuming a linear relationship between capacitance and stroke. 
While simpler, this method provides only an estimation.
Both these approaches involve a trade-off between measurement precision, experimental simplicity and cost. 
Furthermore, they only give an estimate of the stroke.
To address these limitations, an experimental setup was developed to directly measure the piezo stroke using a laser displacement sensor while the piezo is in vacuum and at cryogenic temperature. 
In this paper, the design of the setup and the experimental results are presented.

\subsection{Samples}

\begin{figure*}
	\centering
	\begin{subfigure}{0.18\textwidth}
		\centering
		\includegraphics[width=\linewidth]{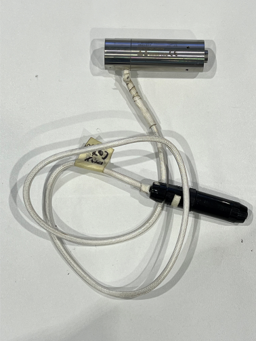}
		\caption{PM piezo}
		\label{fig:PM piezo_photo}
	\end{subfigure}
	\begin{subfigure}{0.25\textwidth}
		\centering
		\includegraphics[width=\linewidth]{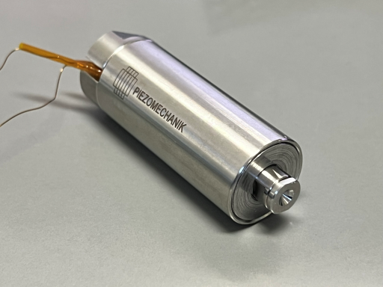}
		\caption{PI piezo}
		\label{fig:PI piezo_photo}
	\end{subfigure}
	\caption{Photo of the two piezo samples which were tested.}\label{fig:piezo}
	
\end{figure*}

We selected two piezos for testing based on previous successful usage in LCLS-II \cite{PI} in the USA and cERL in Japan \cite{PM}.
The LCLS-II piezos are special piezo developed by Physik Instrumente (PI) in collaboration with Fermilab for the operation within the LCLS-II CMs.
The cERL piezo are developed by Piezomechanik (PM) and have shown stable operation at cryogenic temperature.
Since, the operational parameter for both the projects are different from ILC prototype CM, we evaluated the performance of both to identify the one which satisfies our requirements.
\Cref{fig:piezo} shows the two peizos.
Both have two peizo stacks inside the metal casing.
While PM piezo was newly purchased from the vendor, PI piezo was borrowed from Fermilab.
Note that the electrical cable connection to one piezo in PI stack was broken and thus only single stack was tested.
As a first step towards piezo stroke characterization we carried out the capacitance measurement.

\section{Methods}

\subsection{Capacitance Measurement}

\subsubsection{Setup}

\begin{figure*}
	\centering
	
	\begin{subfigure}{0.28\textwidth}
		\centering
		\includegraphics[width=\linewidth]{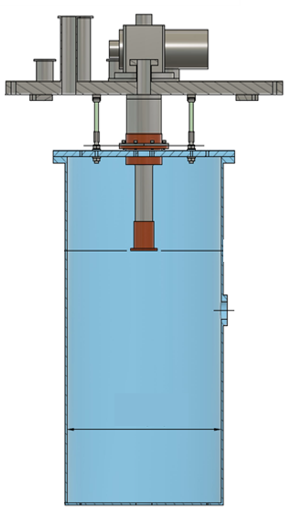}
		\caption{}
		\label{fig:cap_setup_1}
	\end{subfigure}
	\hspace{3mm}   
	\begin{subfigure}{0.28\textwidth}
		\centering
		\includegraphics[width=\linewidth]{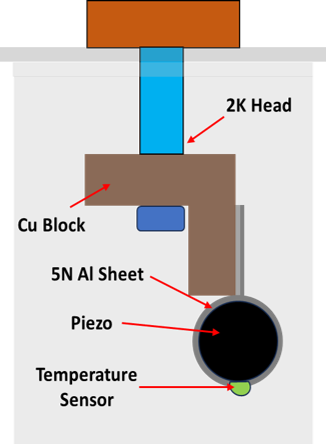}
		\caption{}
		\label{fig:cap_setup_2}
	\end{subfigure}
	\hspace{3mm}   
	\begin{subfigure}{0.35\textwidth}
		\centering
		\includegraphics[width=\linewidth]{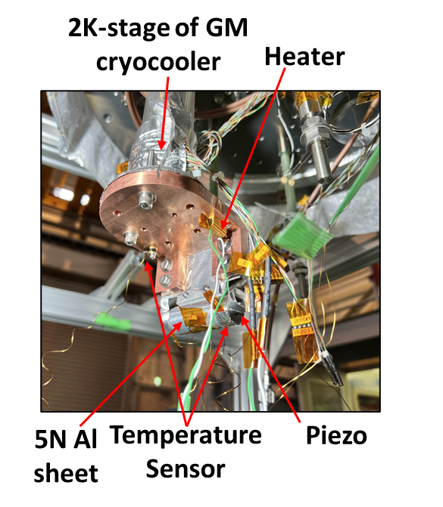}
		\caption{}
		\label{fig:cap_setup_3}
	\end{subfigure}
	
	\caption{The experimental setup for capacitance measurement of two piezos. (a) 2D cross-section of the conduction cooled cryostat used for this measurement. (b) Schematic of the experimental setup. (c) Photo of the setup bolted to the cold-head of the cryostat.}\label{cap_setup}
	
\end{figure*}

The capacitance measurement was carried out in a Gifford-McMahon (GM) cryocooler cooled cryostat, shown in \cref{fig:cap_setup_1}.
The piezo was wrapped in 5N aluminum (99.999\% pure) sheet and bolted onto a copper block.
This copper block was then mounted on to the second stage of the GM cryocooler.
A temperature sensor was fixed on the piezo and another on the copper block.
Additionally, a heater was bolted on to the copper block to speed up warming up the system and for heat load calibration.
\Cref{fig:cap_setup_2,fig:cap_setup_3} shows the schematic layout of the experiment and the actual setup bolted to the cold-head, respectively. 
\subsubsection{Result}

\begin{figure*}
	\centering
	
	\begin{subfigure}{0.4\textwidth}
		\centering
		\includegraphics[width=\linewidth]{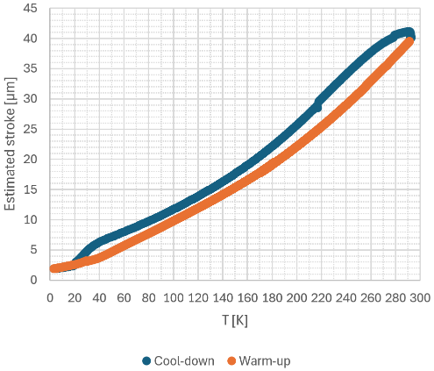}
		\caption{PM piezo}
		\label{fig:cap_result_pm}
	\end{subfigure}
	\hspace{3mm}   
	\begin{subfigure}{0.4\textwidth}
		\centering
		\includegraphics[width=\linewidth]{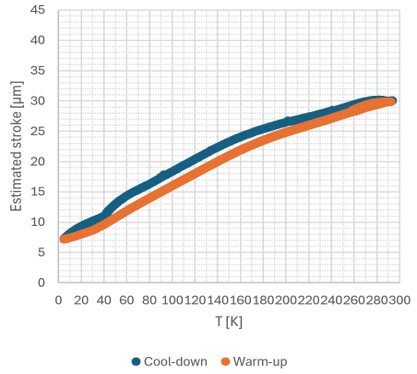}
		\caption{PI piezo}
		\label{fig:cap_result_pi}
	\end{subfigure}
	\hspace{3mm}   
	
	\caption{Estimated drop in piezo stroke with respect to temperature, assuming linear relation between stroke and capacitance drop.}\label{fig:cap_result}
	
\end{figure*}

Both PM and PI piezo went under five cooling and warming up cycles.
There was no degradation in Piezo capacitance after these 5 cycles.
\Cref{fig:cap_result} shows the result for one of the cooling and warm up cycles of each piezo.
The Y-axis of each plot is converted from capacitance to stroke.
Furthermore, the stroke values estimated at 294 and 20 K along with piezo requirements are summarized in \Cref{tab:capcitance}.
For PM piezo, the capacitance and stroke at 294	K was 15 \SI{ }{\micro\farad} and 40 \SI{ }{\micro\meter}, respectively.
After cooling PM piezo down to 20 K the capacitance drops to 0.975 \SI{ }{\micro\farad} which corresponds to a stroke of 2.6 \SI{ }{\micro\meter}.
Similarly, for the PI piezo  the capacitance and stroke at 294	K was 12.4 \SI{ }{\micro\farad} and 30 \SI{ }{\micro\meter}, respectively, which at 20 K that drops to 3.2 \SI{ }{\micro\farad} corresponding to a stroke of 7.7 \SI{ }{\micro\meter}.
Note that since only one stack of PI piezo could be measured the result in \cref{fig:cap_result_pi} and the values of capacitance and stroke written here are scaled by two to estimate the stroke of a properly functioning piezo.

From the capacitance measurement test it is clear that only PI piezo satisfies the stroke requirement for gradient of 40 MV/m, whereas PM piezo only satisfies the requirement for average gradient of 31.5 MV/m.
However, capacitance and piezoelectric coefficients both decrease with temperature but are independent of each other.
Hence, this result is just an estimate of stroke drop.
Therefore, measuring the stroke directly is important.

\begin{table}[]
	\centering
	\caption{Comparison of estimated stroke value of PM and PI piezos at 294 and 20 K. Note that the piezo stroke requirement at 20 K is 2.1 \SI{ }{\micro\meter} and 3.38 \SI{ }{\micro\meter} at gradients of 31.5 MV/m and 40 MV/m, respectively.}
	\label{tab:capcitance}
	
	\begin{tabular}{|c|c|c|}
		\hline
		
		\multirow{2}{*}{\begin{tabular}[c]{@{}c@{}}Temperature\\ {[}K{]}\end{tabular}} 
		& \multicolumn{2}{c|}{Estimated Stroke {[}\si{\micro\meter}{]}} \\ \cline{2-3}
		
		& Piezomechanik & Physik Instrumente \\ \hline
		
		294 & 40 & 30 \\ \hline
		20  & 2.6 & 7.7 \\ \hline
		
	\end{tabular}
\end{table}

\subsection{Direct Stroke Measurement}

\begin{figure}
	\centering
	
	\centering
	\includegraphics[width=0.6\linewidth]{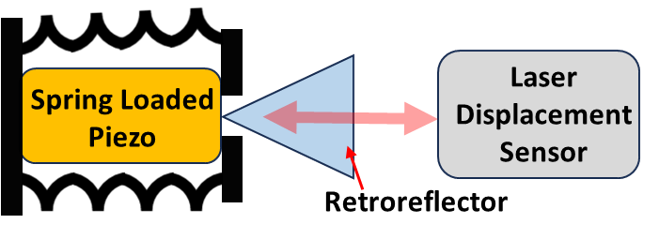}
	
	\caption{Schematic showing the concept of direct stroke measurement setup.}\label{fig:stroke_schematic}
	
\end{figure}

In this section, the details of direct stroke measurement setup is explained.
\Cref{fig:stroke_schematic} shows the conceptual schematic of the measurement setup, the piezo is loaded with a spring to simulate the cavity.
A mirror is mounted in the front and the stroke (displacement) is measured using a commercial, high sensitivity laser displacement sensor.
The details of the actual setup for this measurement is described in \cref{sec:storke_design}.

\subsubsection{Design}\label{sec:storke_design}
\begin{figure*}
	\centering
	\begin{subfigure}{0.4\textwidth}
		\centering
		\includegraphics[width=\linewidth]{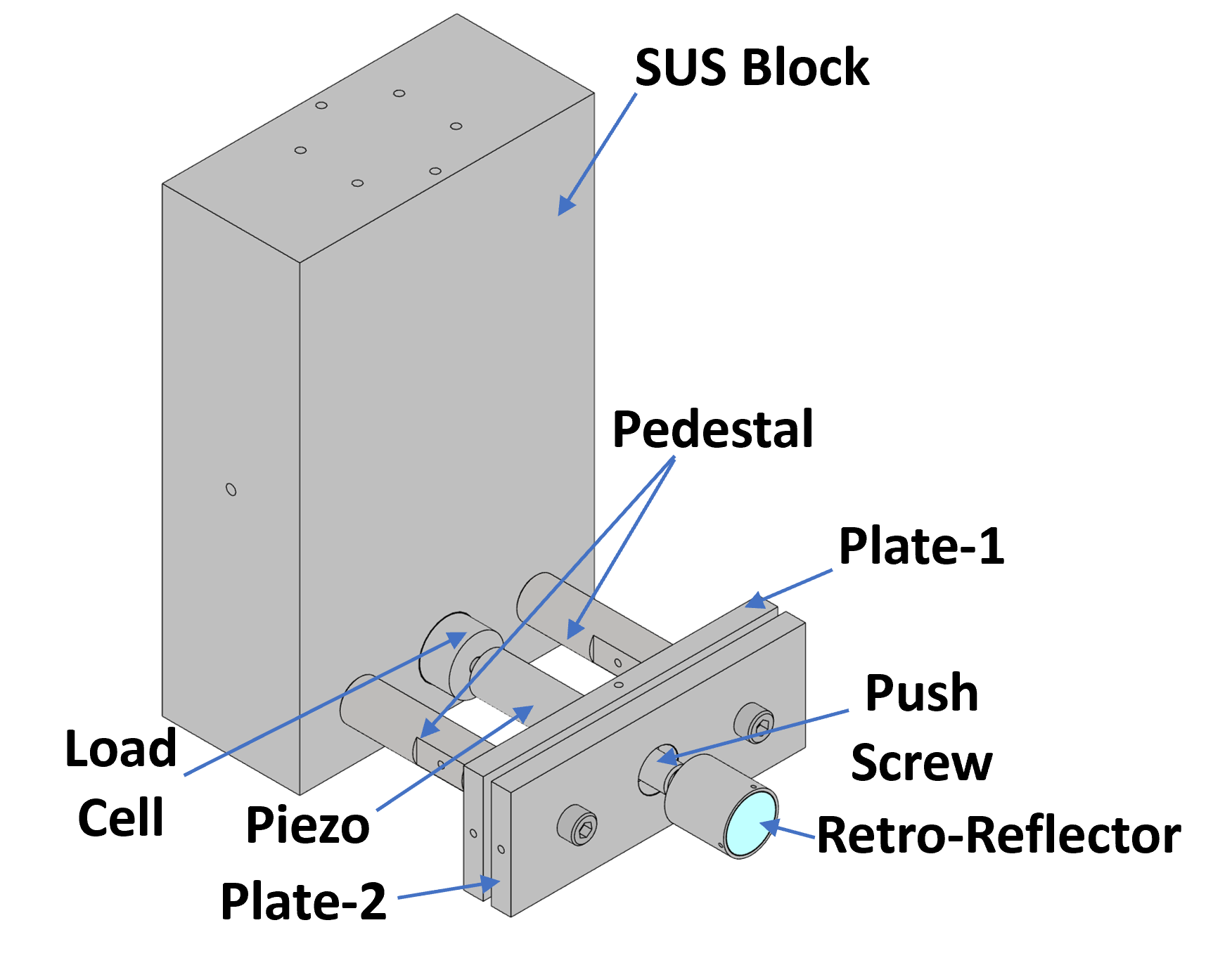}
		\caption{ }
		\label{fig:stroke_setup_2}
	\end{subfigure}
	\hspace{3mm}   
	\begin{subfigure}{0.5\textwidth}
		\centering
		\includegraphics[width=\linewidth]{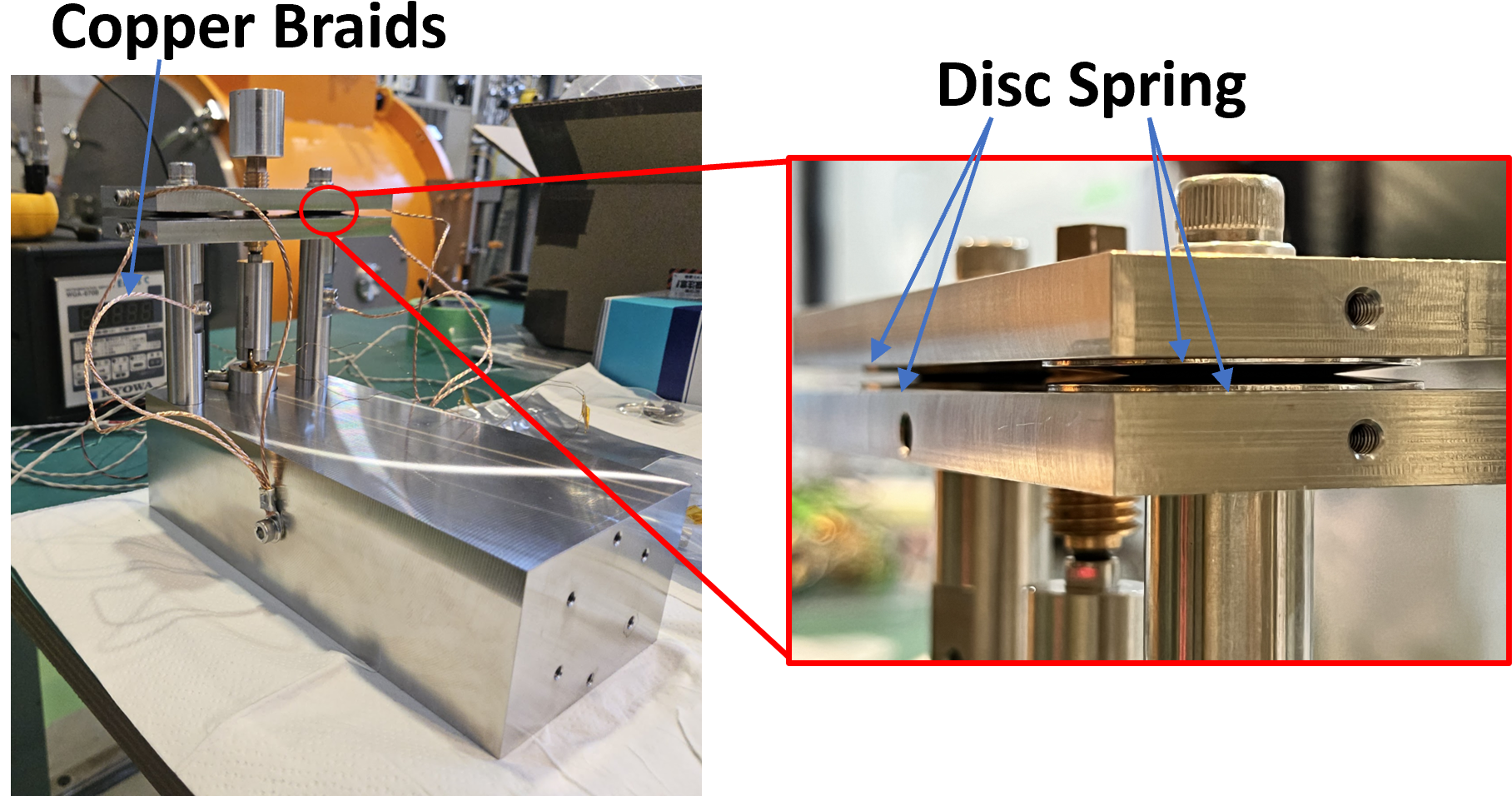}
		\caption{ }
		\label{fig:stroke_setup_4}
	\end{subfigure}
	\hspace{3mm}   
	
	\caption{Part of the direct stroke measurement setup that is housed inside the cryostat. (a) CAD model. (b) Assembled setup outside the cryostat.}\label{fig:stroke_setup}
	
\end{figure*}

\begin{figure*}
	\centering
	
	\centering
	\includegraphics[width=0.5\linewidth]{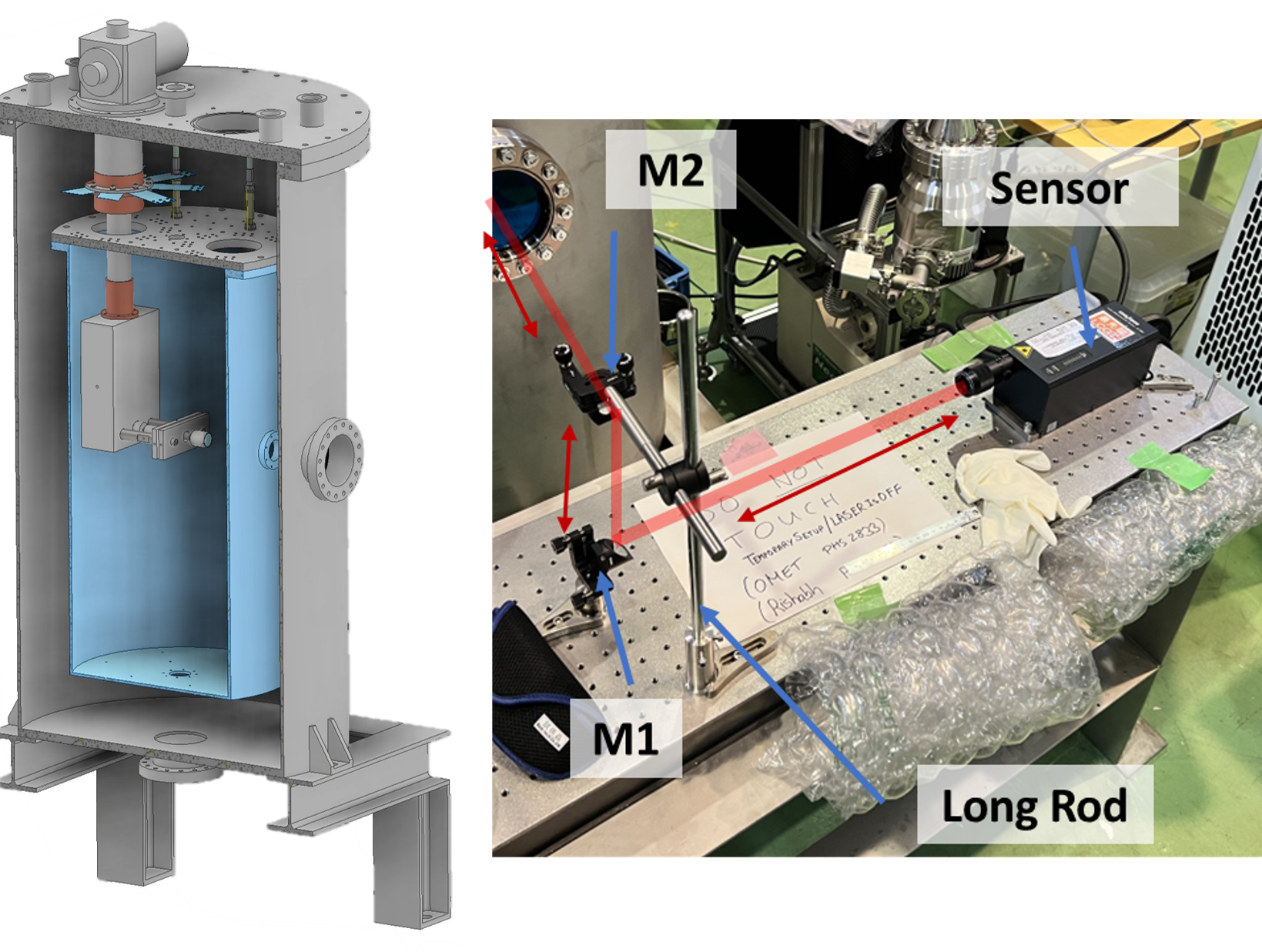}
	
	\caption{Optical layout outside the cryostat along with cross section of CAD assembly of cryostat and stroke measurement setup.}\label{fig:stroke_optical}
	
\end{figure*}

\Cref{fig:stroke_setup,fig:stroke_optical} shows the entire setup of the direct stoke measurement.
The same cryostat as the capacitance measurement experiment was used for the direct stroke measurement.
Since a GM cryocooler is used to cool down this cryostat, the vibration generated from the displacer inside the cold head was a potential noise source as it could be larger than the piezo stroke.
Therefore, we decided to carry out the stroke measurement with the cryocooler turned off. 

\Cref{fig:stroke_setup_2} shows the part of the setup that is cooled down to cryogenic temperature inside the cryostat.
The entire assembly is mounted on a stainless steel (SUS-304) block with dimensions $150 \times 70 \times 235$ mm and a weight of approximately 20 kg. 
Such a large block was selected to slow down the warm-up of the piezo, to ensure sufficient time for carrying out the stroke measurement after the cryocooler is turned off. 
During the capacitance measurement experiment, we observed that the cold head temperature increased at a rate of 1 K/min after turning off the cryocooler, which was not acceptable for the planned test.
Furthermore, the block dampens the vibrations of the setup.
Two steel pedestals (SUS-304) support Steel Plate~1 (SUS-304) on either side from the rear. 
Steel Plate~2 (SUS-304) is secured to the pedestals using two M8 screws, one on each side. 
Two pairs of disc springs are sandwiched between Steel Plates~1 and~2 on both sides, as shown in \Cref{fig:stroke_setup_4} and simulate the cavity.
Disc springs used were because the diameter of the cryostat diameter was not large enough to accommodate coil springs.
Steel Plate~2 has a $\phi 20$ mm through-hole that allows an M14 push screw to pass through the plate and engage with the threaded hole at the center of Steel Plate~1.

The piezo actuator is positioned between a load cell, which rests on the steel block, and the push screw. 
Tightening the push screw compresses the disc springs between the two plates, thereby applying a load to the piezo actuator. 
The applied load is measured using a calibrated load cell.
The load cell was calibrated at 294, 77 and 4 K using a tensile test machine.
A load of \textasciitilde3 kN was applied to both PI and PM piezos during the testing.
A retro-reflector is mounted at the center of the push screw, enabling direct measurement of the piezo stroke.
\Cref{fig:stroke_setup_4} shows the assembled setup, and braided copper wires were connected between the SUS block and the steel plate to ensure uniform and quick cooling of the setup.
\subsubsection{Optical Layout}

The sensor used to carry out the measurement was  a laser displacement sensor developed by Ono Sokki \cite{Ono}.
It works on the principle of laser Doppler vibrometery \cite{vibrometery}.
This device has a maximum measurement range of $\pm5$ m with a resolution of 2.5 nm.
The sensor was housed outside the cryostat on a optical breadboard.
Both the breadboard and the cryostat were fixed to the ground with anchor holes.
The laser from the sensor was injected into the cryostat through an optical window guided by two mirrors (M1 and M2) forming a periscope.
The mirrors were on kinematic mounts to adjust the beam alignment.
A retro-reflector instead of a mirror was used on the push screw since it reflects the light in the incidence direction, simplifying the alignment.
The displacement unit of the sensor was set to a range of 5 \si{\micro\meter}/V, giving us a maximum measurement range of 50 \SI{ }{\micro\meter} and an analog resolution of 1.5 nm.
The details of sensor components and how it works can be found at \cite{Ono}.

\section{Setup Performance}
\subsection{Temperature}
\begin{figure}
	\centering
	
	\centering
	\includegraphics[width=0.8\linewidth]{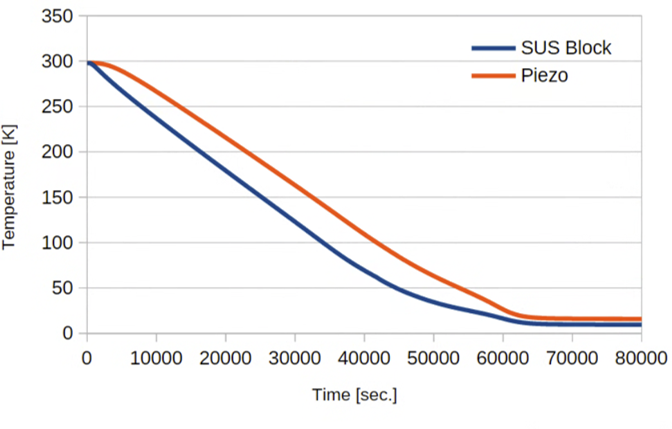}
	
	\caption{Cooling curve of the stroke measurement setup. The total time to reach steady state was 19.5 hours with final temperature of SUS Block and Peizo being 9.5 K and 15.5 K, respectively.}\label{fig:stroke_cooling}
	
\end{figure}

The SUS block was bolted to the cold-head of the cryocooler with indium foil in between.
A temperature sensor was mounted on the SUS block and another on the Piezo.
\Cref{fig:stroke_cooling} shows the cooling curve for this steup.
The total cooling time was around 19.5 hours and final temperature for SUS Block and Piezo were 9.5 K and 15.3 K, respectively.
By making direct thermal contact between Piezo and cold-head lower piezo temperature can be achieved, but for ILC prototype CM testing piezo temperature below 20 K is acceptable.

\subsection{Vibration}
\begin{figure*}
	\centering
	
	\centering
	\includegraphics[width=1.0\linewidth]{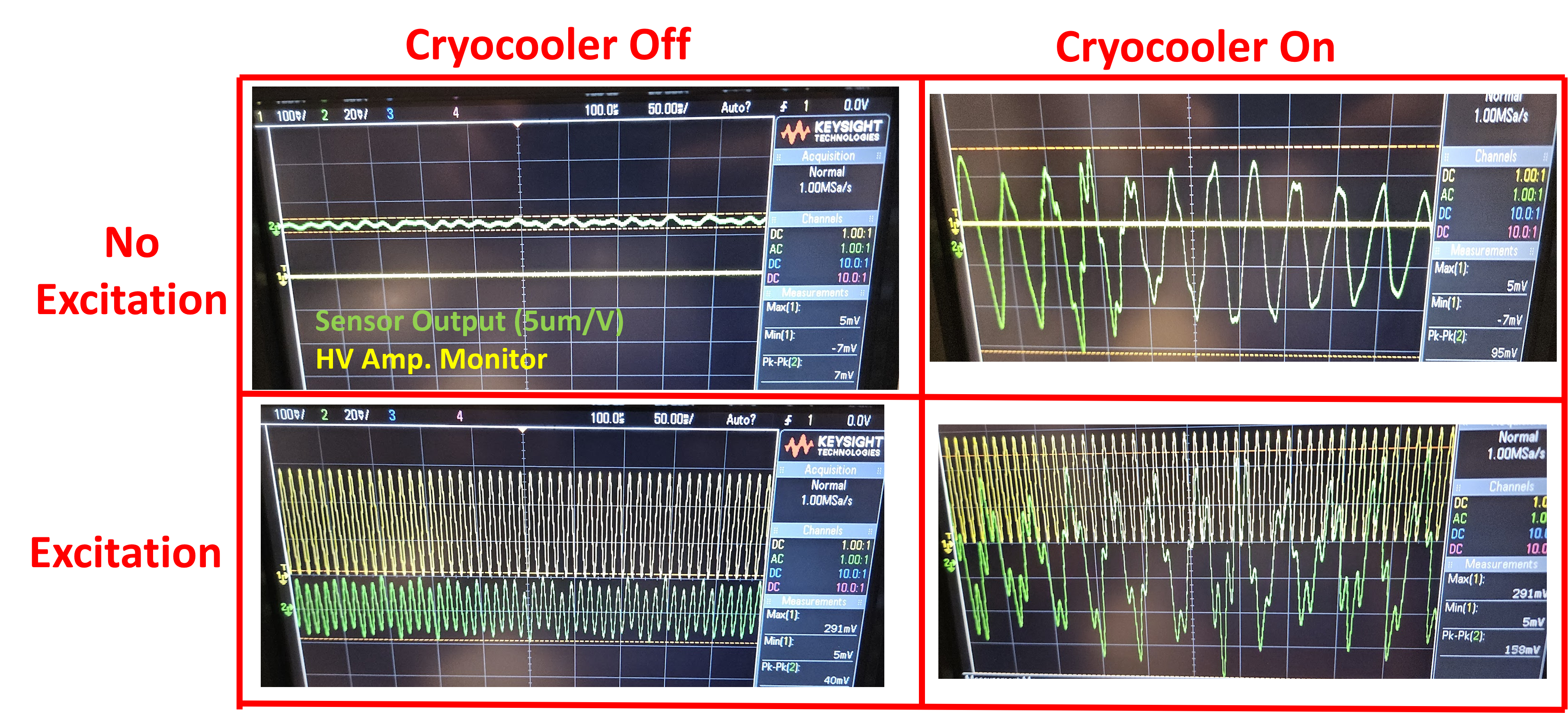}
	
	\caption{Comparison of displacement sensor readout under different operating conditions: cryocooler on/off and with/without piezo excitation. The yellow signal is  HV Amplifier monitor signal while the green signal is the readout of laser displacement sensor.}\label{fig:vib_time}
	
\end{figure*}

\begin{figure*}
	\centering
	
	\centering
	\includegraphics[width=0.7\linewidth]{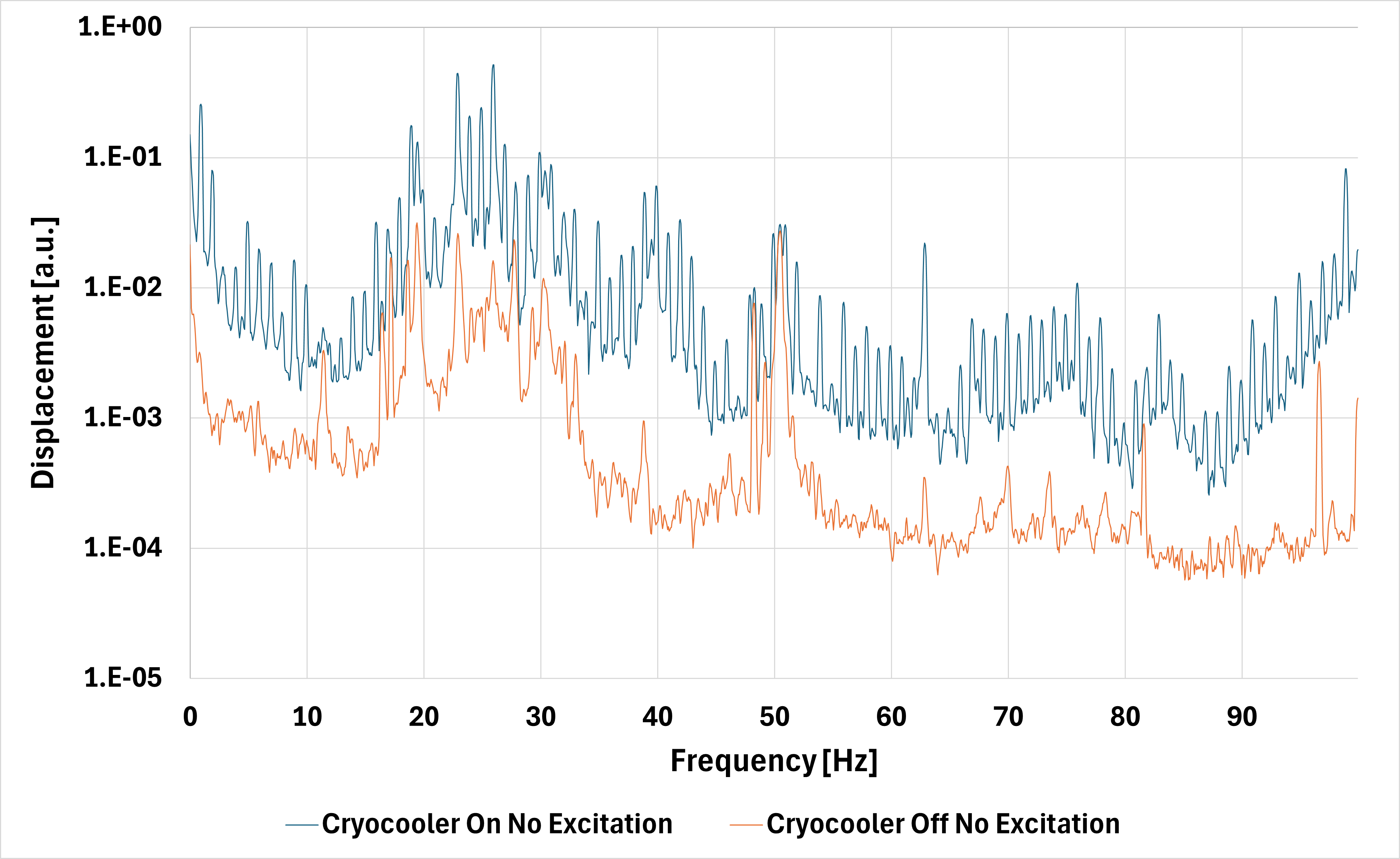}
	
	\caption{Comparison of vibration spectra of the setup with GM cryocooler turned on and off.}\label{fig:vib_fre_1}
	
\end{figure*}

After the cool down was completed the impact of cryocooler vibration on stroke measurement was evaluated in time and frequency domain.
For the time domain analysis, the piezo was excited with a 30 V, 200 Hz half-sine wave signal.
The signal is generated using a \textit{Keysight} signal generator, which is then amplified using a High Voltage (HV) amplifier, before driving the piezo inside the cryostat.
The analog readout from displacement unit of the sensor and monitor port of HV amplifier driving the piezo were monitored on an oscilloscope with cryocooler turned on and off.
\Cref{fig:vib_time} shows the oscilloscope readout.
By comparing the piezo response under excitation with the cryocooler turned on and off, it is evident that the piezo motion can only be clearly observed when the cryocooler is turned off. 
The vibration induced by the cryocooler is significantly larger than the stroke of the piezo actuator, masking the piezo motion when the cryocooler is operating.
Next, the spectra of displacement sensor was monitored with cryocooler turned on and off.
\Cref{fig:vib_fre_1} shows that with the cryocooler turned on the entire setup vibration increases by 1-2 order of magnitude over the entire spectra.
Also, the 1 Hz peak from cryocooler operation are clearly visible.
Therefore, the stroke measurement was carried with cryocooler turned off.

\section{Results}
\begin{figure*}
	\centering
	\begin{subfigure}{0.43\textwidth}
		\centering
		\includegraphics[width=\linewidth]{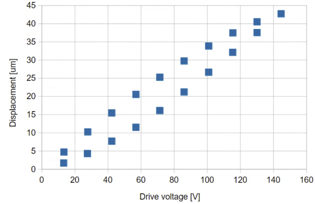}
		\caption{PM piezo at 294 K.}
		\label{fig:PM_294}
	\end{subfigure}
	\hspace{3mm}   
	\begin{subfigure}{0.4\textwidth}
		\centering
		\includegraphics[width=\linewidth]{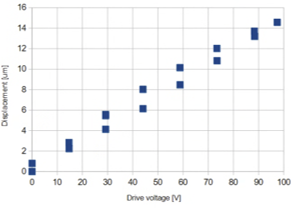}
		\caption{PI piezo at 294 K.}
		\label{fig:PI_294}
	\end{subfigure}
	\hspace{3mm}   
	\begin{subfigure}{0.48\textwidth}
		\centering
		\includegraphics[width=\linewidth]{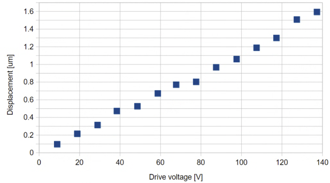}
		\caption{PM piezo at 20 K.}
		\label{fig:PM_20}
	\end{subfigure}
	\hspace{3mm}   
	\begin{subfigure}{0.4\textwidth}
		\centering
		\includegraphics[width=\linewidth]{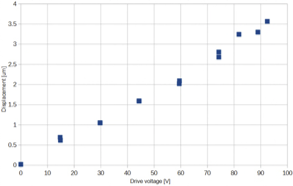}
		\caption{PI piezo at 20 K.}
		\label{fig:PI_20}
	\end{subfigure}
	\hspace{3mm}   
	
	\caption{The results of direct stroke mesaurement for PM piezo and PI piezo at 294 K and 20 K. Note that the result for PI piezo is for single stack.}\label{fig:stroke_results}
	
\end{figure*}

Both Piezo A and B were individually tested.
The stroke of each peizo was measured under DC excitation.
For PM piezo the maximum rated voltage is 150 V and hence it was tested until 140 V.
Whereas for PI piezo the maximum rated voltage is 100 V and it was tested until 95 V.
The results are shown in \cref{fig:stroke_results}.
The stroke of PM piezo was measured to be 43 \SI{ }{\micro\meter} at 294 K and drops by 96.3\% to 1.6 \SI{ }{\micro\meter} at 20 K.
Whereas, for PI piezo the stroke at 294 K was measured to be 14.5 \SI{ }{\micro\meter} and drops by 73.8\% to 3.8 \SI{ }{\micro\meter} at 20 K.
Note that the result for PI piezo is for a single stack, for a fully operational piezo the stroke at 294 K and 20 K will be 29 \SI{ }{\micro\meter} and 7.8 \SI{ }{\micro\meter}, respectively.

From the direct stroke measurement, it is clear that only the PI piezo satisfies the stroke requirement for the ILC prototype cryomodule. 
Furthermore, the actual stroke measured for the PM piezo is 1.6 \SI{ }{\micro\meter}, which is 61.5\% smaller than the estimated stroke of 2.6 \SI{ }{\micro\meter} obtained from the capacitance measurement. 
This indicates that capacitance-based stroke estimation can give incorrect results and may not be reliable.

\section{Conclusion}
Current methods for measuring the strokes of piezos used in SRF cavity tuners either use indirect measurements with the cavity, which are accurate but slow and expensive, or use capacitance measurements, which are easy to perform but only give rough estimates of the stroke.
In this work, we developed and applied a new setup to measure the piezo stroke directly using a laser displacement sensor inside a cryocooler-cryostat.
This setup allows the stroke to be measured directly, accurately, and at low cost.
The setup was used to test two piezo actuators for the ILC prototype cryomodule and to select the appropriate unit.
We believe that this method will be useful for reliable quality control in large-scale projects that use many SRF cavities.

\section{Declaration of competing interest}

The authors declare that they have no known competing financial interests or personal relationships that could have appeared to influence the work reported in this paper

\section{Acknowledgments}

This work was supported by the MEXT Development of key element technologies to improve the performance of future accelerators Program Japan Grant Number JPMXP1423812204. 
The authors would like to thank Dr. Tomohiro Yamada of the High Energy Accelerator Research Organization for his support in this experiment.
The authors also thank Prof. Takayuki Tomaru of the National Astronomical Observatory of Japan for lending the laser displacement sensor used for the stroke measurements.
We also thank the Mechanical Engineering Center for machining the components used in the stroke and capacitance measurement experiments.
We are grateful to the Cryogenic Science Center at KEK for allowing us to use the tensile testing machine to calibrate the load cell at cryogenic temperatures.

\section{Data Availability}

Data will be made available on request.


\end{document}